\documentclass[conference]{IEEEtran}
\IEEEoverridecommandlockouts
\usepackage{cite}
\usepackage{amsmath,amssymb,amsfonts}
\usepackage{algorithmic}
\usepackage{graphicx}
\usepackage{textcomp}
\usepackage{xcolor}
\usepackage{comment}
\usepackage{tikz, pgfplots}
\pgfplotsset{compat=1.18}
\def\BibTeX{{\rm B\kern-.05em{\sc i\kern-.025em b}\kern-.08em
    T\kern-.1667em\lower.7ex\hbox{E}\kern-.125emX}}

\makeatletter
\newcommand{\linebreakand}{%
  \end{@IEEEauthorhalign}
  \hfill\mbox{}\par
  \mbox{}\hfill\begin{@IEEEauthorhalign}
}
\makeatother

\title{Multilevel Coset Codes on Lattices\\

\thanks{This research was supported in part by the U.S. Department of Commerce’s National Telecommunications and Information Administration (NTIA) under the Public Wireless Supply Chain Innovation Fund Grant Program (Award 24-60-IF2415: ASPEN - Advanced Signal Processing Enhancement for Next-Generation Open Radio Units), administered by the National Institute of Standards and Technology.}
}

\author{
\IEEEauthorblockN{Leopold Bertholet, Chloe Makdad, Stephen Mackes, Daniel Chew, Matthew Robinson}
\IEEEauthorblockA{\textit{Rampart Communications}\\
Linthicum Heights, MD, USA \\
\{lbertholet, cmakdad, stephen, dchew, matt\}@rampartcommunications.com}
}

\begin{document}

\maketitle

\begin{abstract}
This work introduces coset Bombe codes, a novel class of multilevel coset codes that generalize polar codes to dense lattice structures. By leveraging multilevel coding with non-binary codes designed for the lattice modulations and making use of Voronoi shaping, Bombe codes integrate the geometric strengths of dense lattices such as $D_4$ with the capacity-approaching properties of polar codes. Experimental results in additive white Gaussian noise (AWGN) channels demonstrate that coset Bombe codes significantly outperform both BICM and MLC state-of-the-art schemes on 16-QAM. The proposed scheme simulated on AWGN achieves up to 0.8 dB of gain and reduces block size latency by half while maintaining superior bit and block error rate (BER/BLER) performance on codewords of 256 and 1024 bits.
\end{abstract}

\section{Introduction and Background}

\subsection{Lattices in Communication Systems}

\IEEEPARstart{M}{odern} digital communication traces back to Shannon's foundational work \cite{shannonnoise}, which established that reliable communication is possible below channel capacity. For the additive white Gaussian noise (AWGN) channel, while a Gaussian input distribution is optimal, practical systems utilize finite constellations. Successive developments in Turbo \cite{turbo}, low density parity check (LDPC) \cite{gallager1963low}, \cite{mackay97near}, and polar codes \cite{Arikan2009PolarCodes} have approached the AWGN channel capacity, often by combining binary forward error correction (FEC) with traditional formats like quadrature-amplitude modulation (QAM).

However, these results are primarily asymptotic, and asymptotic results do not immediately yield system design. Modern low-latency requirements for reliable communication at small and medium block sizes make large binary codewords impractical, rendering the structural properties of modulation a critical performance bottleneck. Consequently, dense lattices \cite{conway1999sphere} have emerged as an integrated framework to maximize distance between points (coding gain) and minimize average power (shaping gain) \cite{latticecapacityzamir}. While shaping gain is asymptotically capped at $1.53$ dB \cite{forney_multidimensional}, both lattice and binary coding gains effectively increase Euclidean distance to achieve capacity \cite{latticecapacityzamir},\cite{calderbank}.

The evolution of coded modulation includes strategies like trellis-coded modulation (TCM), which uses lower bits protected by a convolutional code as subset selectors for uncoded higher bits in a modulation map \cite{ungerboeck1}. Multilevel coding (MLC) was proposed contemporaneously and imposes separate codes and rates on different bit levels of a modulation \cite{imaimultilevel}. Multilevel schemes are especially powerful in the context of multistage decoding (MSD), in which levels are decoded in serial order and codes on higher levels are not used in decoding lower levels. Despite its apparent suboptimality relative to maximum-likelihood decoding, MSD achieves capacity under appropriate selection of level rates \cite{wachsmannMLC}. TCM and MLC using MSD thus share a serial decoding process that exploits redundancy on lower levels of a modulation scheme to achieve higher code rates on later levels. Forney specifically formulated TCM in terms of coset codes, in which the sublattice structure of a lattice quotient constellation is used to construct levels for coding \cite{forney1988coset}. In a more general framework, multilevel coset codes use chains of nested sublattices in a lattice constellation to create levels for an MLC scheme, allowing separate codes with appropriate rates to be applied to each level \cite{forney2000sphere}. In particular, polar codes can be used in conjunction with MLC to achieve capacity \cite{seidl2013polar} and a combined polar-coded multilevel construction may be viewed in terms of the lattice partition framework \cite{liu2018construction}. 

Bit-interleaved coded modulation (BICM) schemes are generally more flexible than MLC and can closely approximate channel capacity \cite{ciareBICM}. BICM currently serves as the industry benchmark, alongside recent shaping methods, including geometric shaping \cite{mirani2020low}, \cite{li2025coded} and probabilistic shaping \cite{Bocherer2015PAS}, \cite{SchulteBocherer2016CCDM}. Despite this progress, simultaneously achieving shaping, lattice, and FEC gains with low complexity soft-decision decoding in high dimensions remains a persistent challenge.

In this paper, we introduce \textit{Bombe codes}\footnote{Spelled like Turing's cryptanalytic machine from WW2, pronounced ``Bahm-buh'' like Rejewski's Polish ancestor.}, a generalization of polar codes to dense lattices. These codes achieve shaping, lattice, and FEC gains while maintaining a complexity profile suitable for practical systems. Specifically, we define \textit{coset} Bombe codes as multilevel codes that use polar codes on each bit significance level following the framework by Forney in \cite{forney1988coset}, \cite{forney2000sphere}, and decode them with multistage decoding.

\subsection{Multilevel Codes}

We introduce the set partitioning construction and multistage decoder for multilevel codes and omit any more general framework. Assume a modulation that assigns more than 1 bit per real dimension of a signal space, such as the I/Q plane. We refer to each bit position as a \textit{level}, with each level being protected by its own FEC. Under AWGN and multistage decoding, performance is optimized by Ungerboeck's set partition construction \cite{ungerboeck1}, which increases minimum distance from one code level to the next. The multistage decoding process operates serially in bit significance order: the estimated bit values in the first level determine how the next level is demodulated, so a decision must be made on lower bits before proceeding \cite{imaimultilevel}.

FEC is applied in each level separately. That is, in the $i^{\text{th}}$ level $k_b^{(i)}$ bits are collected and encoded with code $C^{(i)}$ to obtain a codeword of length $n_b^{(i)}$. The resulting coded modulation is capacity achieving as long as the component codes each achieve capacity and are chosen with rates $R^{(i)}$ that approximate the channel capacities of the level constellations \cite{wachsmannMLC}. In contrast, BICM schemes are not guaranteed to achieve capacity, although they may approximate it closely \cite{BICM}. Notably, the use of polar codes as component codes within a multilevel scheme allows for alternative code construction techniques, such as estimating the capacity or Bhattacharyya parameters of synthetic channels after decoding \cite{seidl2013polar}, \cite{vangala2015comparative}. When we simulate polar codes in this work, we use such constructions, which closely approximate the aforementioned physical channel capacities.

\subsection{Outline}

Section \ref{sec:codesoncosets} develops coset Bombe codes in detail. We set up the framework at a high level, focusing on lattice partition chains and the resulting quotient groups. Section \ref{sec:codedesign} covers code construction for generalized polar codes over $(\mathbb{Z}/2\mathbb{Z})^d$. In Section \ref{sec:encodingsection}, we outline the multilevel code construction using the lattice partition chain. Section \ref{sec:Demod} details how to implement the multistage decoder, and Section \ref{sec:decodingsection} explains our decoding scheme. The computational complexity of the decoder is examined in Section \ref{sec:complexity}. Finally, the specific experimental setup, configuration parameters, and primary results are outlined in Section \ref{sec:resultssection}, where we measure the coding gain of our coset Bombe codes against the state-of-the-art polar coded modulations in the AWGN channel.

\section{Codes on Cosets}
\label{sec:codesoncosets}

Coset Bombe codes extend classical multilevel coding on a QAM by replacing bit levels with levels of a lattice quotient constellation. The lattice modulation map is due to  \cite{conway2003fast}, while the multilevel construction is described by a lattice partition chain after the manner of \cite{forney2000sphere}.

Let $\Lambda \subset \mathbb{R}^d$ be a full rank lattice with $d\times d$ generator matrix $M$. For an integer $r \geq 2$, the quotient $\Lambda/r\Lambda$ has $r^d$ elements which are the cosets of $r\Lambda$ in $\Lambda$. The constellation is formed by mapping $x \in (\mathbb{Z}/r\mathbb{Z})^d$ to $y = M x \in \Lambda/r\Lambda$. Additionally, we shift $\Lambda$ by a small displacement such that each coset of $r\Lambda$ has a unique representative in the Voronoi cell of $r\Lambda$ \cite{conway2003fast}.

Given a lattice $\Lambda_0$, an \textit{$L$-level lattice partition chain} is a finite sequence of nested sublattices
\begin{align}
    \Lambda_0 \supset \Lambda_1 \supset \cdots \supset \Lambda_L \notag
\end{align}
that decomposes the quotient $\Lambda_0/\Lambda_L$ into successive quotients
\begin{align}
    Q_i = \Lambda_{i}/\Lambda_{i+1}. \notag 
\end{align}
Each $y \in \Lambda_0/\Lambda_L$ is then uniquely specified by its sequence of coset representatives:
\begin{align}
    y = q^{(0)} + q^{(1)} + \cdots + q^{(L-1)}, \qquad q^{(i)} \in Q_i. \notag
\end{align}
Thus, an $L$-level lattice partition chain induces an $L$-level structure in the modulation map. The $i^\text{th}$ level encodes $\log_2(|Q_i|)$ bits, potentially using a non-binary code as necessary.

We will focus on quotients of the form $\Lambda/r\Lambda$, where $r = 2^s$, with $s$ being the number of bit significance levels. Consider a partition chain obtained by geometric scaling:
\begin{align}
    \Lambda \supset 2^{a_1}\Lambda \supset 2^{a_2}\Lambda \supset \cdots \supset r\Lambda, \notag
\end{align}
where $0=a_0 <a_1 < \cdots < a_L=s$. The $i^\text{th}$ quotient is 
\begin{align}
    Q_i = 2^{a_i}\Lambda / 2^{a_{i+1}} \Lambda \cong (\mathbb{Z}/2^{s_i}\mathbb{Z})^d, \notag
\end{align}
where $s_i = a_{i+1} - a_i$ and $\cong$ denotes a group isomorphism. For the remainder of this paper, we use the following unit step chain where $a_i = i$, with $L=s$ levels:
\begin{align}
    \Lambda \supset 2\Lambda \supset 4 \Lambda \supset \cdots \supset r\Lambda. \notag
\end{align}
For every level $i$ we have
\begin{align}
    Q_i = 2^i\Lambda/2^{i+1}\Lambda \cong \Lambda/2\Lambda, \notag
\end{align}
so this partition chain has the property that every quotient group has $2^d$ points and is isomorphic to $(\mathbb{Z}/2\mathbb{Z})^d$. Each level has minimum distance equal to the minimum distance of $2^{i+1}\Lambda$, and for every $0 \leq i \leq L-2$ these minimum distances satisfy
\begin{align}
    d_\text{min}(Q_{i+1}) = 2d_\text{min}(Q_{i}) \notag
\end{align}
so that the minimum distance doubles at every stage. This property holds regardless of the choice of generator matrix $M$ for $\Lambda$.

Applying an MLC to the unit step chain assigns a separate code to each level $y^{(i)} \in 2^{i}\Lambda/2^{i+1}\Lambda$. At each stage we employ a non-binary coded modulation over $(\mathbb{Z}/2\mathbb{Z})^d$, or $\text{GF}(2^d)$, matching the additive structure of the quotient group. While the code alphabet remains invariant across stages, the code rates are optimized to exploit the doubling of the minimum distance $d_{\text{min}}$.

In the scalar case $\Lambda = \mathbb{Z}$, this framework recovers classical MLC on $2^s$-ary uniformly spaced pulse amplitude modulation constellations. The partition chain $\mathbb{Z}/2\mathbb{Z}/4\mathbb{Z}/\cdots/r\mathbb{Z}$ induces a labeling where each $y \in \{0,1,\ldots,r-1\}$ is identified by its canonical binary expansion. This coincides with Ungerboeck's set partitioning strategy, where stage $i$ protects bit $i$ of the expansion \cite{ungerboeck1}. Under multistage decoding \cite{wachsmannMLC}, the least significant bit (LSB) $y^{(0)}$ is therefore decoded first, followed by bits of increasing significance.

For general $d$-dimensional lattices, the decomposition of $y \in \Lambda / r\Lambda$ is most efficiently represented via the coefficient vector $x \in (\mathbb{Z}/r\mathbb{Z})^d$. The linearity of the map $y = Mx$ ensures that the lattice partition chain induces an equivalent partition on the coefficient space:
\begin{align}
    \mathbb{Z}^d/2\mathbb{Z}^d/4\mathbb{Z}^d/\cdots/r\mathbb{Z}^d. \notag
\end{align}
Consequently, the vector $x$ admits a unique bit-level decomposition 
\begin{align}
    x = x^{(0)} + 2x^{(1)} + \cdots + 2^{L-1} x^{(L-1)}, \;\; x^{(i)} \in (\mathbb{Z}/2\mathbb{Z})^d, \notag
\end{align}
analogous to the unique coset-level decomposition
\begin{align}
    y = y^{(0)} + 2y^{(1)} + \cdots + 2^{L-1} y^{(L-1)},\;\; y^{(i)} \in \Lambda/2\Lambda. \notag
\end{align}
A multistage decoder thus processes the binary vectors $x^{(i)}$ in sequence. In each step $i$, the decoder jointly resolves the $d$ bits corresponding to the $i^\text{th}$ significance level across all lattice dimensions.

\subsection{Code Design}
\label{sec:codedesign}

As mentioned in the previous section, we choose the levels of our constellation to be the quotients $2^i\Lambda/2^{i+1}\Lambda$. Each quotient is isomorphic as a group to $(\mathbb{Z}/2\mathbb{Z})^d$ or the additive group of $\text{GF}(2^d)$. There is a large class of FECs over $\text{GF}(2^d)$ that achieve capacity. We choose a polar code with the same binary encoder matrix as in \cite{Arikan2009PolarCodes} without any elements other than $0$ and $1$ so that only the additive structure, $(\mathbb{Z}/2\mathbb{Z})^d$, is used. This convention allows for $d$ parallel binary polar encodings at the transmitter, although the lattice structure will require joint decoding of dimensions at the receiver.

For code construction, we adopt a unified reliability ranking across all bit positions, yielding code construction within each level in addition to rate allocation across levels \cite{seidl2013polar}. While multilevel schemes often rely on the asymptotic capacity rule for rate allocation across levels \cite{wachsmannMLC}, polar codes offer greater flexibility by enabling reliability sequences that encompass all bit-level codes. In our implementation, bit capacities in all levels are estimated via Monte Carlo simulation of rate $0$ successive cancellation across a range of SNR values. This approach is particularly advantageous for polar codes over finite fields; by permitting bit-level freezing within $\text{GF}(2^d)$ symbols, it provides the finer control necessary for the optimal construction of Bombe codes. Moreover, the capacity rate allocation rule is asymptotic, whereas the numerical polar reliability estimations fully account for the finite block size behavior.

Regarding the code construction for the simulations in this work, we restrict our results to the $D_4$ lattice to bound complexity and only consider $r=4$. This is throughput-equivalent to a $16$-QAM as it transmits $4$ bits per I/Q channel use. Rather than presenting the full reliability sequence, which is fairly long, Table \ref{tab:RateTable}
shows the LSB code rate corresponding to each overall code rate for our bit block size $n_b=1024$ simulations. For comparison, we also include the rates that the capacity rule would assign to each level to show that they are in close agreement. The most significant bit rate is readily calculated since the rates have average equal to total rate.

\begin{table}[]
    \centering
    \scriptsize
    \begin{tabular}{|c||c|c|c|}
    \hline
        Total Rate & $7/16$ & $11/16$ & $15/16$ \\
        \hline \hline
        Numerical LSB Rate & 0.066 & 0.385 &  0.875 \\
        \hline
        Capacity Rule LSB Rate  & 0.094 & 0.445 &  0.879  \\   \hline
    \end{tabular}
    \vspace{.5em}
    \caption{Table of code rates for LSB in the two level scheme presented in this work. Actual rates are for $n_b=1024$}
    \label{tab:RateTable}
    \vspace{-3em}
\end{table}

\subsection{Encoding}
\label{sec:encodingsection}

As discussed above and following the standard multilevel approach, each level is encoded with a non-binary polar code over $\text{GF}(2^d)$ operating on $d$-bit symbols \cite{finitefieldpolar}. Because we are only using the additive structure of $\text{GF}(2^d)$ at the transmitter, this is equivalent to $d$ parallel binary polar encoders. Within a single level, we encode each of the parallel polar codes with different code construction, determined as in Section \ref{sec:codedesign}.

\begin{figure*}[!t]
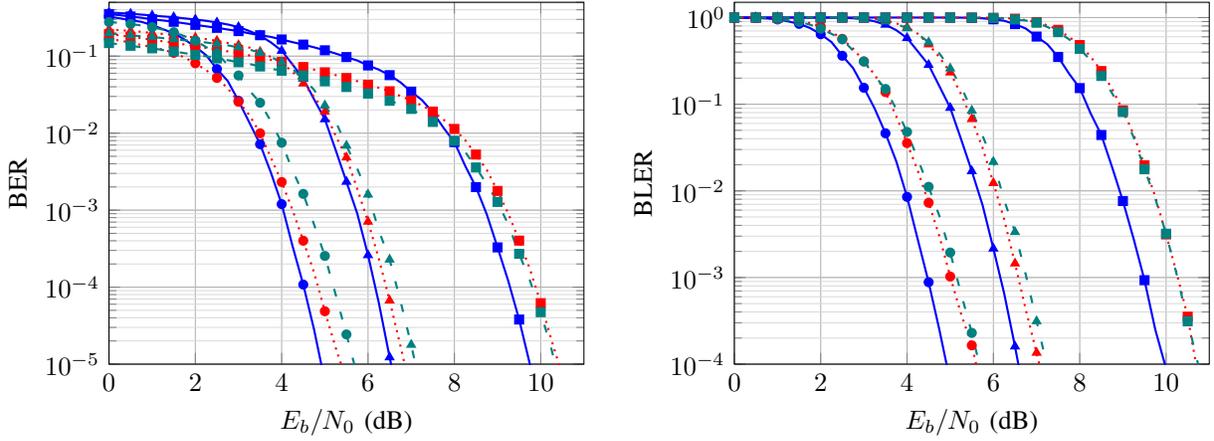

    \centering
    \include{ber256}
    \caption{BER (left) and BLER (right) curves for $D_4$ Bombe codes compared to state-of-the-art polar coded modulations, all with $n_b = 256$. We use $r=4$ for our Bombe code and a $16$-QAM for the polar codes. See the legend in Fig. \ref{fig:1024}.}
    \label{fig:256}
\end{figure*}

\begin{figure*}[!t]
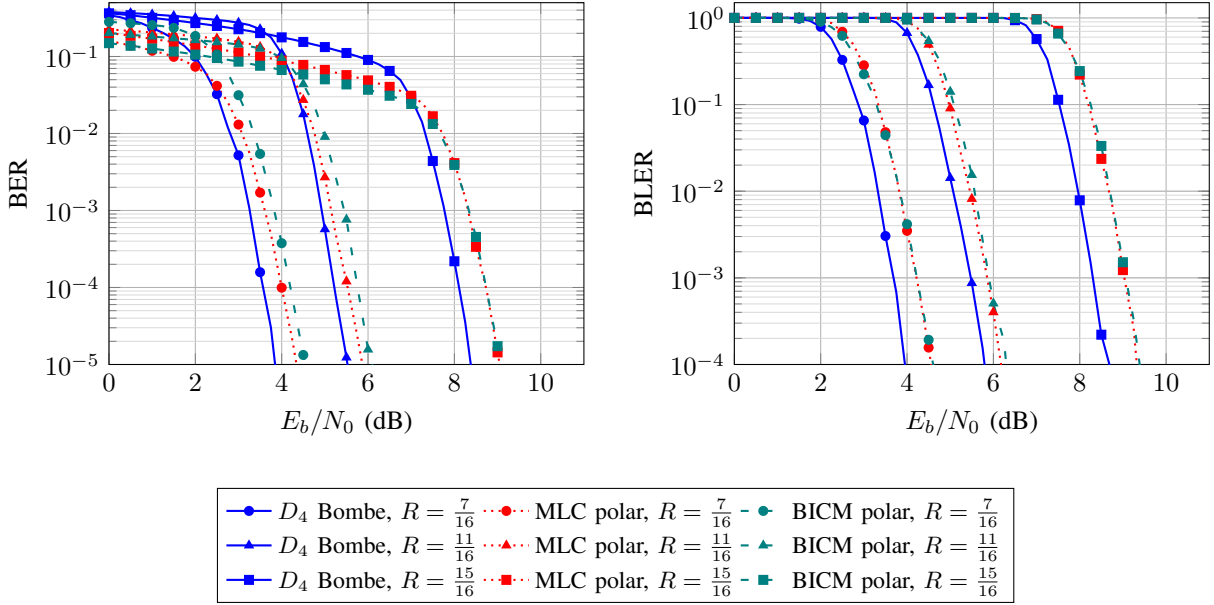

    \centering
    \include{ber1024}
    \caption{BER (left) and BLER (right) curves for $D_4$ Bombe codes compared to state-of-the-art polar coded modulations, all with $n_b = 1024$. We use $r=4$ for our Bombe code and a $16$-QAM for the polar codes.}
    \label{fig:1024}
\end{figure*}

Each level produces a codeword of size $n_b^{(i)} = n_l d\quad \text{bits},$ where $n_l$ is the number of lattice points per codeword. Level $i$ encodes $k^{(i)}_b$ bits per codeword and operates at rate $R^{(i)} = k^{(i)}_b/n_b^{(i)}$. The overall FEC coding rate as specified in our results (Table \ref{tab:RateTable} and Figs. \ref{fig:256} and \ref{fig:1024}) is $R = k_b / n_b$, where $k_b = \sum_{i=0}^{s-1}k_b^{(i)}$ is the total number of information bits per codeword across all $s$ levels and $n_b=\sum_{i=0}^{s-1}n_b^{(i)}$ is the total number of encoded bits per codeword. The overall coding rate in bits/dimension is then given by
\begin{align*}
    R^{(0)} + R^{(1)} + \cdots + R^{(s-1)} = sR.
\end{align*}

The specific encoder used in each level and dimension is a systematic polar code with distinct rate and code design in each level \cite{arikansystematic}, \cite{sarkis2015flexible}, \cite{morozovsystematic}. Furthermore, we concatenate with a CRC to enable CRC-aided succesive cancellation list (CA-SCL) decoding \cite{talvardylistdecoding}, \cite{crcaidniuchen}.

\subsection{Demodulation}
\label{sec:Demod}

Assume the data of an AWGN noisy lattice symbol $\mu \in \mathbb{R}^d$ and some shaped lattice constellation $\mathcal{C} \subset \Lambda \subset \mathbb{R}^d$. We use the probability distribution on $\mathcal{C}$ that is proportional to 
\begin{align}
    p_\sigma (x | \mu) = \exp \left( -\dfrac{1}{2\sigma^2} \left\| x - \mu \right\|^2 \right) \label{eq:demod_p}
\end{align}
for $x \in \mathcal{C}$, where $\sigma^2$ is the noise variance. This probability mass function (PMF) can then be marginalized to obtain the PMF on the quotient $\Lambda/2\Lambda$. Then upon obtaining an estimate $h^{(0)}$ for $\Lambda/2\Lambda$, we can condition and marginalize $p$ to obtain the PMF on $h^{(0)} + 2\Lambda/4\Lambda$. Similarly, given an estimate $h^{(i-2)}$ of the first $i-1$ bit levels $0, \ldots, i-2$, we can condition on the first $i-1$ bit levels and marginalize the bits of levels higher than $i-1$ to obtain the PMF on $h^{(i-2)} + 2^{i-1} \Lambda/2^i \Lambda$. By subtracting $h^{(i-2)}$ we view this PMF as if it were on the quotient $2^{i-1} \Lambda /2^i \Lambda$.  This level $i$ PMF conditioned on the decoded value $h^{(i-2)} \in \Lambda/2^{i-1} \Lambda$ (where $h^{(-1)}=0$) is now defined by 

\begin{align}
    p^{(i)}_\sigma(x | \mu, h^{(i-1)}) &= \sum_{y \in \mathcal{C}^{(i)}_{x}} p_\sigma (y+h^{(i-1)}  | \mu) \label{eq:marginalize_p}
\end{align}
for $x \in 2^{i-1}\Lambda/2^i \Lambda$, where $\mathcal{C}^{(i)}_{x} = \{y \in \mathcal{C} \mid y = x \mod 2^{i} \Lambda\}$. The complexity of the marginalization step may be simplified using max-log approximations. 

In our results, we use $\Lambda = D_4$ and $r=4$, so $| \mathcal{C}| = 256$ and $|\Lambda/2\Lambda| = 16$.

\subsection{Decoding}
\label{sec:decodingsection}

At the receiver, the data is demodulated with \eqref{eq:demod_p} and then decoded with an $s$-stage multistage decoder. The soft information passed into each stage is conditioned on the decoded values of all less significant bit levels according to \eqref{eq:marginalize_p}. Each component is a non-binary polar code over $\text{GF}(2^d)$ decoded with CA-SCL. To improve the performance of MSD with list decoding, list scores and branches are propagated across component decoders \cite{zhang2021path}; the branch with correct CRC having the lowest path metric is selected after the final decoding stage.

\subsection{Complexity} \label{sec:complexity}

We now consider the demodulation and decoding complexity of coset Bombe codes. The demodulation map of \eqref{eq:demod_p} produces a data structure storing the probability of every lattice point in the signal constellation $\mathcal{C} = \Lambda/r\Lambda$. Because $\mathcal{C}$ is $d$-dimensional and non-orthogonal, by default this map requires the distance between the received $\mu$ and every possible lattice point $x$ in the constellation to be computed. Thus, the demodulation map is an $\mathcal{O}(r^d)$ computation for each of the $n_l$ received lattice points per codeword. Similarly, the marginalization map of \eqref{eq:marginalize_p} is carried out before each decoding stage and produces an array of $2^d$ probabilities. Letting $r=2^s$, there are $s$ decoding stages under MSD. At each stage $i=0, 1, \ldots, s-1$, the current soft information for each lattice point is marginalized down from $2^{(s-i)d}$ to $2^d$ probabilities, requiring $\mathcal{O}(2^{(s-i-1)d})$ additions.

The standard polar decoding complexity is $\mathcal{O}(n\log n)$ for codeword length $n$ \cite{Arikan2009PolarCodes}. For polar codes over finite fields $\text{GF}(2^d)$, the complexities $C_U$ and $C_L$ of the upper and lower channel splitting functions are nontrivial and may depend asymptotically on $d$. As such, the complexity of the $s$-stage multilevel coset decoding process is
\begin{align*}
    C_l = \mathcal{O}(s\cdot n_l\log(n_l)\cdot (C_U+C_L)).
\end{align*}
A comparable multilevel polar code on a QAM with the same spectral efficiency and number of bits per codeword would have an $s$-stage decoding process with $n_p = d\cdot n_l$ bits at each level, resulting in decoding complexity
\begin{align}
    C_p = \mathcal{O}(s\cdot n_p\log(n_p)) = \mathcal{O}(s\cdot dn_l\log(dn_l)). \label{eq:mlc_complexity}
\end{align}
For $d>1$, the size of the polar encoding matrix on each bit level in coset Bombe is $n_l$ and compares favorably to $dn_l$ for multilevel polar in (\ref{eq:mlc_complexity}). However, in general the polar channel splitting functions, $C_U$ and $C_L$, are inherently more complex over $\text{GF}(2^d)$. In our implementation, the two functions require element-wise multiplication and multidimensional cyclic convolution of PMF arrays with $2^d$ entries, involving $\mathcal{O}(2^d)$ and $\mathcal{O}(2^d\log(2^d))$ operations, respectively \cite{yuan2018construction}.

Thus, the computational complexity depends exponentially on the lattice dimension $d$ in both the demodulation and decoding steps. Specifically, the implementations of lattice point demodulation and of upper and lower functions in polar codes over $\text{GF}(2^d)$ are key in determining the efficiency of the overall scheme. While the current implementation complexity may be reasonable for small-dimensional lattices like $D_4$, extension of our work to higher dimensions will be the subject of future work.

\section{Results}
\label{sec:resultssection}

We evaluate the performance of our multilevel coset Bombe codes against existing state-of-the-art polar coded modulations, namely BICM polar and MLC polar, using bit error rate (BER) and block error rate (BLER). All decoders are CA-SCL with list size $8$ and use the CRC dictated by the 5G standard for each given rate and block size \cite{multiplexing20185g}. We consider bit block sizes $n_b = 256$ and $1024$, both of which are used by the 5G control channel, at a variety of rates. As noted above, we limit our results to the $D_4$ lattice with $r=4$ for our Bombe codes, resulting in $s=2$ coding levels. To ensure a fair comparison in terms of spectral efficiency, our BICM polar and MLC polar results use a $16$-QAM modulation scheme.

Results are seen in Fig. \ref{fig:256} and Fig. \ref{fig:1024}, which share the legend in Fig \ref{fig:1024}. The type of code is indicated by color and line style, and rate indicated by line marker. The performance of codes with the same rate results in those curves being grouped together. 

Across all examined rates and block sizes, our Bombe codes see gain over MLC and BICM polar in both BER and BLER. The multilevel framework of MLC polar provides gains in BER over BICM polar, most pronounced at smaller block size. The coset Bombe codes seen here also benefit from being multilevel. However, unlike the gains of MLC polar over BICM, coset Bombe gains over BICM also arise from shaping and are preserved when moving from small to large block size. At $n_b=1024$, $R=15/16$, we observe gains of $0.71\text{ dB}$ in BER and $0.77 \text{ dB}$ in BLER over MLC polar. This is a a result of a combination of shaping and coding gain resulting from use of the $D_4$ lattice. MLC polar gains over BICM in BER are also almost fully erased when considering BLER instead, while Bombe still clearly outperforms both state-of-the-art polar coded modulations.

Under AWGN, we also observe that the use of a coset Bombe code with $r=4$ can achieve the same or better bit and block error rate performance than an $n_b=1024$, $R = 3/4$ BICM Polar code modulated to a $16$-QAM with half the bit block size. In other words, the use of a coset Bombe code with this configuration can reduce block size latency by half. 

\section{Discussion and Future Work}

We have introduced coset Bombe codes, a novel class of multilevel coset codes that generalizes previous work on multistage lattice coding. Our proposed scheme combines shaping, lattice, and FEC gain into a single framework, providing up to $0.8$ dB of gain across a range of realistic scenarios. 

Directions for future work on Bombe codes include extensions to higher-throughput constellations (e.g., analogues of $64$-QAM and $256$-QAM) and larger lattices (e.g., $E_8$). In particular, generalization to larger lattices may require improvements in the asymptotic computational complexity of demodulation and decoding. Other improvements in complexity, such as optimization of runtime and latency, as well as opportunities for hardware integration, are also of interest. Finally, while this work has focused on combining multilevel coset codes with generalized polar codes, the approach presented here may be extended to other code families (e.g., LDPC).

\bibliographystyle{IEEEtran}
\bibliography{bibliography}

@article{li2025coded,
  title={Coded modulation schemes for Voronoi constellations},
  author={Li, Shen and Mirani, Ali and Karlsson, Magnus and Agrell, Erik},
  journal={IEEE Transactions on Communications},
  year={2025},
  publisher={IEEE}
}

@article{forney1988coset,
  title={Coset codes. I. Introduction and geometrical classification},
  author={Forney, G David},
  journal={IEEE Transactions on Information Theory},
  volume={34},
  number={5},
  pages={1123--1151},
  year={1988},
  publisher={IEEE}
}

@ARTICLE{shannonnoise,
  author={Shannon, C.E.},
  journal={Proceedings of the IRE}, 
  title={Communication in the Presence of Noise}, 
  year={1949},
  volume={37},
  number={1},
  pages={10-21},
  doi={10.1109/JRPROC.1949.232969}}

@INPROCEEDINGS{turbo,
  author={Berrou, C. and Glavieux, A. and Thitimajshima, P.},
  booktitle={Proceedings of ICC '93 - IEEE International Conference on Communications}, 
  title={Near Shannon limit error-correcting coding and decoding: Turbo-codes. 1}, 
  year={1993},
  volume={2},
  number={},
  pages={1064-1070 vol.2},
  doi={10.1109/ICC.1993.397441}}

@book{gallager1963low,
  title={Low-Density Parity-Check Codes},
  author={Gallager, Robert G},
  year={1963},
  publisher={MIT Press},
  address={Cambridge, MA}
}

@inproceedings{mackay97near,
  author = {David J. C. MacKay and Radford M. Neal},
  title = {Near {S}hannon limit performance of low density parity check codes},
  booktitle = {Electronics Letters},
  volume = {33},
  number = {6},
  pages = {457--458},
  year = {1997}
}

@article{Arikan2009PolarCodes,
  author  = {Ar{\i}kan, Erdal},
  title   = {Channel polarization: A method for constructing capacity-achieving codes for symmetric binary-input memoryless channels},
  journal = {IEEE Transactions on Information Theory},
  volume  = {55},
  number  = {7},
  pages   = {3051--3073},
  year    = {2009},
  month   = jul,
  doi     = {10.1109/TIT.2009.2021379}
}

@ARTICLE{wachsmannMLC,
  author={Wachsmann, U. and Fischer, R.F.H. and Huber, J.B.},
  journal={IEEE Transactions on Information Theory}, 
  title={Multilevel codes: theoretical concepts and practical design rules}, 
  year={1999},
  volume={45},
  number={5},
  pages={1361-1391},
  doi={10.1109/18.771140}}

@ARTICLE{latticecapacityzamir,
  author={Erez, U. and Zamir, R.},
  journal={IEEE Transactions on Information Theory}, 
  title={Achieving 1/2 log (1+SNR) on the AWGN channel with lattice encoding and decoding}, 
  year={2004},
  volume={50},
  number={10},
  pages={2293-2314},
  doi={10.1109/TIT.2004.834787}}

@ARTICLE{forney_multidimensional,
  author={Forney, G.D. and Wei, L.-F.},
  journal={IEEE Journal on Selected Areas in Communications}, 
  title={Multidimensional constellations. I. Introduction, figures of merit, and generalized cross constellations}, 
  year={1989},
  volume={7},
  number={6},
  pages={877-892},
  doi={10.1109/49.29611}}

@book{conway1999sphere,
  author    = {Conway, John H. and Sloane, Neil J. A.},
  title     = {Sphere Packings, Lattices and Groups},
  edition   = {3rd},
  publisher = {Springer},
  year      = {1999},
  series    = {Grundlehren der mathematischen Wissenschaften},
  volume    = {290},
  isbn      = {978-0-387-98585-5},
  doi       = {10.1007/978-1-4757-6568-7}
}

@ARTICLE{calderbank,
  author={Calderbank, A.R.},
  journal={IEEE Transactions on Information Theory}, 
  title={The art of signaling: fifty years of coding theory}, 
  year={1998},
  volume={44},
  number={6},
  pages={2561-2595},
  doi={10.1109/18.720549}}

@ARTICLE{ungerboeck1,
  author={Ungerboeck, G.},
  journal={IEEE Transactions on Information Theory}, 
  title={Channel coding with multilevel/phase signals}, 
  year={1982},
  volume={28},
  number={1},
  pages={55-67},
  doi={10.1109/TIT.1982.1056454}}

@ARTICLE{imaimultilevel,
  author={Imai, H. and Hirakawa, S.},
  journal={IEEE Transactions on Information Theory}, 
  title={A new multilevel coding method using error-correcting codes}, 
  year={1977},
  volume={23},
  number={3},
  pages={371-377},
  doi={10.1109/TIT.1977.1055718}}

@INPROCEEDINGS{ciareBICM,
  author={Caire, G. and Taricco, G. and Biglieri, E.},
  booktitle={Proceedings of IEEE International Symposium on Information Theory}, 
  title={Bit-interleaved coded modulation}, 
  year={1997},
  volume={},
  number={},
  pages={96-},
  doi={10.1109/ISIT.1997.613011}}

@article{mirani2020low,
  title={Low-complexity geometric shaping},
  author={Mirani, Ali and Agrell, Erik and Karlsson, Magnus},
  journal={Journal of Lightwave Technology},
  volume={39},
  number={2},
  pages={363--371},
  year={2020},
  publisher={IEEE}
}

@article{conway2003fast,
  title={Fast quantizing and decoding and algorithms for lattice quantizers and codes},
  author={Conway, John and Sloane, Neil},
  journal={IEEE Transactions on Information Theory},
  volume={28},
  number={2},
  pages={227--232},
  year={2003},
  publisher={IEEE}
}

@misc{vangala2015comparative,
  title         = {A Comparative Study of Polar Code Constructions for the AWGN Channel},
  author        = {Harish Vangala and Emanuele Viterbo and Yi Hong},
  year          = {2015},
  eprint        = {1501.02473},
  archivePrefix = {arXiv},
  primaryClass  = {cs.IT},
  doi           = {10.48550/arXiv.1501.02473},
  url           = {https://arxiv.org/abs/1501.02473}
}

@ARTICLE{arikansystematic,
  author={Arikan, Erdal},
  journal={IEEE Communications Letters}, 
  title={Systematic Polar Coding}, 
  year={2011},
  volume={15},
  number={8},
  pages={860-862},
  doi={10.1109/LCOMM.2011.061611.110862}}

@misc{sarkis2015flexible,
  title         = {Flexible and Low-Complexity Encoding and Decoding of Systematic Polar Codes},
  author        = {Sarkis, Ghassan and Tal, Ido and Giard, Pascal and Vardy, Alexander and Thibeault, Claude and Gross, Warren J.},
  year          = {2015},
  eprint        = {1507.03614},
  archivePrefix = {arXiv},
  primaryClass  = {cs.IT},
  url           = {https://arxiv.org/abs/1507.03614}
}

@ARTICLE{morozovsystematic,
  author={Morozov, Ruslan and Trifonov, Peter},
  journal={IEEE Wireless Communications Letters}, 
  title={Successive and Two-Stage Systematic Encoding of Polar Subcodes}, 
  year={2019},
  volume={8},
  number={3},
  pages={877-880},
  doi={10.1109/LWC.2019.2898195}}

@ARTICLE{talvardylistdecoding,
  author={Tal, Ido and Vardy, Alexander},
  journal={IEEE Transactions on Information Theory}, 
  title={List Decoding of Polar Codes}, 
  year={2015},
  volume={61},
  number={5},
  pages={2213-2226},
  doi={10.1109/TIT.2015.2410251}}

@ARTICLE{crcaidniuchen,
  author={Niu, Kai and Chen, Kai},
  journal={IEEE Communications Letters}, 
  title={CRC-Aided Decoding of Polar Codes}, 
  year={2012},
  volume={16},
  number={10},
  pages={1668-1671},
  doi={10.1109/LCOMM.2012.090312.121501}}

@ARTICLE{BICM,
  author={Caire, G. and Taricco, G. and Biglieri, E.},
  journal={IEEE Transactions on Information Theory}, 
  title={Bit-interleaved coded modulation}, 
  year={1998},
  volume={44},
  number={3},
  pages={927-946},
  doi={10.1109/18.669123}}

@inproceedings{zhang2021path,
  title={Path metric inherited SCL decoding of multilevel polar-coded systems},
  author={Zhang, Dexin and Wu, Bolin and Niu, Kai},
  booktitle={2021 IEEE wireless communications and networking conference workshops (WCNCW)},
  pages={1--6},
  year={2021},
  organization={IEEE}
}

@article{finitefieldpolar,
  title={Source and Channel Polarization Over Finite Fields and Reed--Solomon Matrices},
  author={Mori, Ryuhei and Tanaka, Toshiyuki},
  journal={IEEE Transactions on Information Theory},
  volume={60},
  number={5},
  pages={2720--2736},
  year={2014},
  publisher={IEEE},
  doi={10.1109/TIT.2014.2307572}
}

@techreport{multiplexing20185g,
    author = "{3GPP}",
    title = "{NR; Multiplexing and channel coding}",
    institution = "3rd Generation Partnership Project (3GPP)",
    series = "TS",
    number = "38.212",
    version = "16.5.0"
}

@article{seidl2013polar,
  title={Polar-coded modulation},
  author={Seidl, Mathis and Schenk, Andreas and Stierstorfer, Clemens and Huber, Johannes B},
  journal={IEEE Transactions on Communications},
  volume={61},
  number={10},
  pages={4108--4119},
  year={2013},
  publisher={IEEE}
}

@article{Bocherer2015PAS,
  author  = {B{\"o}cherer, Georg and Steiner, Fabian and Schulte, Patrick},
  title   = {Bandwidth Efficient and Rate-Matched {LDPC} Coded Modulation with Probabilistic Shaping},
  journal = {IEEE Transactions on Communications},
  year    = {2015},
  volume  = {63},
  number  = {12},
  pages   = {4651--4665},
  doi     = {10.1109/TCOMM.2015.2485616}
}

@article{SchulteBocherer2016CCDM,
  author  = {Schulte, Patrick and B{\"o}cherer, Georg},
  title   = {Constant Composition Distribution Matching},
  journal = {IEEE Transactions on Information Theory},
  year    = {2016},
  volume  = {62},
  number  = {1},
  pages   = {430--434},
  doi     = {10.1109/TIT.2015.2503982}
}

@inproceedings{yuan2018construction,
  title={Construction and decoding algorithms for polar codes based on 2$\times$ 2 non-binary kernels},
  author={Yuan, Peihong and Steiner, Fabian},
  booktitle={2018 IEEE 10th International Symposium on Turbo Codes \& Iterative Information Processing (ISTC)},
  pages={1--5},
  year={2018},
  organization={IEEE}
}

@article{forney2000sphere,
  title={Sphere-bound-achieving coset codes and multilevel coset codes},
  author={Forney, G David and Trott, Mitchell D and Chung, Sae-Young},
  journal={IEEE Transactions on Information Theory},
  volume={46},
  number={3},
  pages={820--850},
  year={2000},
  publisher={IEEE}
}

@article{liu2018construction,
  title={Construction of capacity-achieving lattice codes: Polar lattices},
  author={Liu, Ling and Yan, Yanfei and Ling, Cong and Wu, Xiaofu},
  journal={IEEE Transactions on Communications},
  volume={67},
  number={2},
  pages={915--928},
  year={2018},
  publisher={IEEE}
}

\end{document}